# Optimization parameter design for proton irradiation accelerator

AN Yu-Wen (安宇文)*[1,2], JI Hong-Fei(纪红飞)[3], WANG Sheng (王生)[1,2], XU Shou-Yan(许守彦)[1,2]

[1]*China Spallation Neutron Source (CSNS), Institute of High Energy Physics (IHEP), Chinese Academy of Sciences (CAS)，Dongguan 523803, China*
[2]*Dongguan Institute of Neutron Science (DINS), Dongguan 523808, China*
[3]*Institute of High Energy Physics (IHEP), Beijing 100049, China*

The proton irradiation accelerator is widely founded for industry application, and should be designed as compact, reliable, and easy operate. A 10 MeV proton beam is designed to be injected into the slow circulation ring with the repetition rate of 0.5 Hz for accumulation and acceleration, and then the beam with the energy of 300MeV will be slowly extracted by third order resonance method. For getting a higher intensity and more uniform beam, the height of the injection bump is carefully optimised during the injection period. Besides, in order to make the extracted beam with a more uniform distribution, a RF Knock-out method is adopted, and the RF kicker's amplitude is well optimised.

## 1 Introduction

The basic design of a compact proton accelerator system for irradiation accelerator is described. The system consists of a 50KeV H⁻ ion source, a 10 MeV tandem accelerator, and a 300MeV slow cycling synchrotron. When H⁻ beam travels along the tandem, the electrons of the H⁻ are scraped through Argon gas in low pressure, and then the proton beam is transported through beam line. The DC mode proton beam is injected into the ring through a two-bump structure with the beam current of 100$u$A. When the energy of the proton beam reaches 300MeV, the beam will be slowly extracted by third order resonance method. The main parameters of the irradiation proton accelerators are listed in Table 1 [1].

Table 1: Main parameters of irradiation proton accelerator

| Parameters | Units | Values |
| --- | --- | --- |
| Ring circumference | m | 33.6 |
| Inj. Energy | MeV | 10 |
| Ext. Energy | MeV | 300 |
| Nominal Tunes(H/V) in ring | | 1.70/1.20 |
| Repetition rate | Hz | 0.5 |
| Beam emmittance in ring (99%) | $\pi$ mm-mrad | 25/25 |
| Beam emmittance for one inj. pulse (99%) | $\pi$ mm-mrad | 6.83 |
| Proton Number in ring | | $5*10^9$ |
| Momentum deviation | | $5*10^{-4}$ |
| Inj. scheme | | Multi-turn injection |
| Ext. scheme | | Slow extraction (RFKO) |

Third order resonance method is a common technique for better control of the extracted beam characters. When the horizontal tune of the beam is near $n \pm 1/3$ (n is a random integer), particles far away for the beam core may become unstable, that is because the sextuples in the ring establish a limited stable area in the shape of triangle. In order to control the extracted beam more precisely, a RF knock-out method is adopted to make the particles amplitude growth beyond the triangle. The RF kicker's amplitude is carefully modified during the period of the extraction to make the extraction beam more flatten.

## 2 Optimisation of the injection bump height decrease pattern

The Lattice of the proton irradiation synchrotron is designed with two super-period, and each one super-period consists of two FODO structure. 8 dipoles with 45 degrees and 12 quadruples distributed separately along the ring, and the layout of the synchrotron is shown in figure 1 [2]. Figure 2 shows the lattice parameter [3] of the synchrotron. Long drift with non-dispersion is 9.2m, and is reserved for the installation of the RF cavity, injection elements, extraction elements, bumps, sextuple and so on. Four sextuples, which two of them are located in the dispersion free region to produce the driving term for slow extraction, and the other two of them, are located in the dispersion region to compensate the chromaticity. Although the natural horizontal chromaticity of the synchrotron is -1.2, the chromaticity still needs to be compensated for a better performance of the RF kicker.

---

*Corresponding author (email: anyw@ihep.ac.cn)

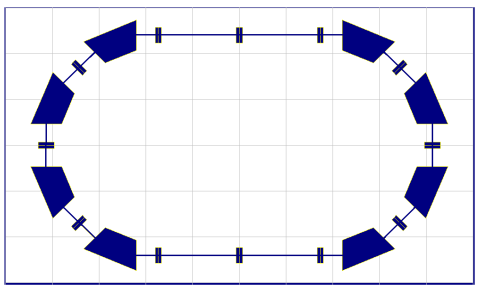

**Fig. 1.** The layout of the irradiation proton synchrotron.

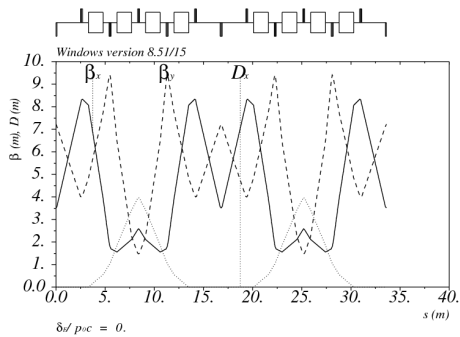

**Fig. 2.** The lattice parameter of the irradiation proton synchrotron.

In our design, the H⁻ beam changes to proton in tandem accelerator, so for getting a more efficient injection, multi-turn injection method is adopted, and two kickers with phase advance π located symmetrically about the injection point. Figure 3 shows the locations of the two kickers and the initial bump height [4]. The red and blue line represent horizontal and vertical closed orbit respectively.

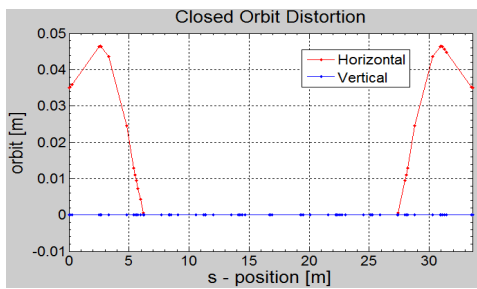

**Fig. 3.** The location of the kicker and the initial bump.

Figure 4 is the sketch of the multi-injection in horizontal phase space. The red line represents the septum, which is a special magnet to separate the injection beam and circulating beam. When proton beam from transport line is injected into the synchrotron, two kickers whose phase advance is π form a bump and the injection beam is accepted by the circulating beam. And then the strengths of those two kickers are adjusted to make the bump decreasing to fill the hollow of the circulating beam. The thickness of the septum is a major source for beam loss, unfortunately, it is impossible to make that more and more thinner. So we emphasize study the decreeing pattern of the bump height to get more uniform beam distribution and more particles circulating in the ring.

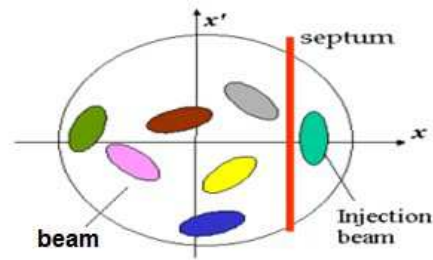

**Fig. 4.** The sketch of the multi-turn injection. The red line represents the septum and the colorful elliptic represents the injection beam or circulating beam.

In our simulation, the thickness of septum is 0.1mm, and the septum locates off center orbit 12mm, and the root mean square of the width of the injection beam is 5mm. Firstly, according to the physical aperture (0.035m) and twiss parameter of the synchrotron, the acceptance $\varepsilon_{acc}$ of the ring should be figured. And then, the maximum turn, $n = \varepsilon_{acc} / (1.5 * \varepsilon_{rms})$ [5], is computed through the relationship between injection beam emmittance $\varepsilon_{rms}$ and acceptance of the synchrotron. At last, the maximum value of injection turn is figured, and it equals 14. In order to get better injection efficiency, lots of injection conditions were changed in our simulation, including the injection beam size (or injection beam distribution) and decreasing pattern of the bump height. Finally, we found that mismatch injection that means the beta function of injection beam is different with beta function of the synchrotron may lead more particles accumulation. The decreasing patterns of the bump height between square root and curve we designed are carefully compared according to the particles survival in the synchrotron, and we found that the decreasing pattern of the bump height we designed is better than square pattern. Figure 5 shows the results of the comparison between fitting pattern and square pattern, and after the circulating beam revolutions 40 turns, more particles are survival. Figure 6 shows beam distribution in phase space on 40[th] turn corresponds to the fitting curve, and beam distribution is almost uniform in phase space.

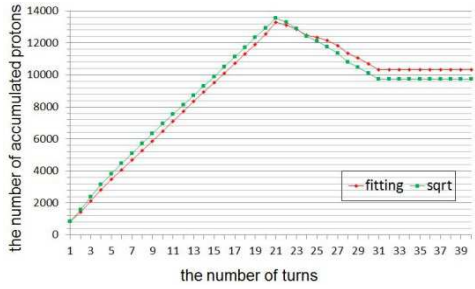

**Fig. 5.** The comparison between fitting pattern and the square pattern of the bump height.

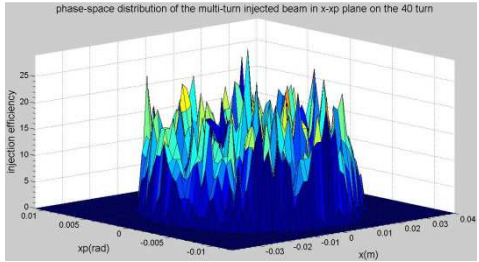

**Fig. 6.** Distribution of the circulating beam on 40$^{th}$ turn in phase space.

## 3  Optimisation the slow extraction

### a)  Planning sextuple families for beam extraction

Either half integer or third-integer resonance can be used for slow extraction, but the current trend is towards using third-order resonance for more controllable spills [6]. It is difficult to use fourth order resonance because the separatrices are becoming too small [7]. In third order resonance applications, the phase space is shaped into triangle by sextuples, and only particles in the triangle are stable, but with the effects of write noise, RF kick, and those particles may dilute in the inner of the triangle, and become unstable. The upper part of Figure 7 shows 9 particles with different initial position revolution 100 turns. And from the figure, the particles in the outer space become unstable and the amplitude grow bigger and bigger, however, the other particles still keep stable in the inner area. For describing the third-order resonance more precisely, the distance h (depicted in the bottom of Fig. 7) between the upright separatrix and the upright axis (X') is conveniently introduced, where $h = \frac{2}{3}\frac{\varepsilon}{S} = \frac{4\pi}{S}\delta Q$, and $\varepsilon$ or $\delta Q$ is tune distance between particles and third-order resonance, and S is the normalized sextuple strength [6]. In our simulation, two sextuples were placed in the synchrotron, and one is located in the third period in the dispersion area, the other is located in the second period in the dispersion free area. Firstly, we adjusted the chromatic sextuples and make the horizontal chromaticity equal -0.0006, and value of the normalised sextuples strength is -24.0826 m$^{-1/2}$. And then the harmonic sextuple was placed in the second region, and the phase advance between two sextuples is $\pi/3$. By using Accelerator Toolbox [4], we can get distance $h$, and the comparison between theoretical calculation and the numerical tracking is listed in Table 2. From the table, the distance $h$ of theoretical computation is somewhat bigger than that of numerical tracking. And that is easy to understand, there is a little error in getting value from tracking because the separatrix is a thin bar not just a line. For considering saving the sextuple strength, two sextuples whose phase advance is $\pi/3$ are enough to extract the spill from the synchrotron, because the overall strengths of the sextuples equal the plus value of the two individual sextuples with special phase advance.

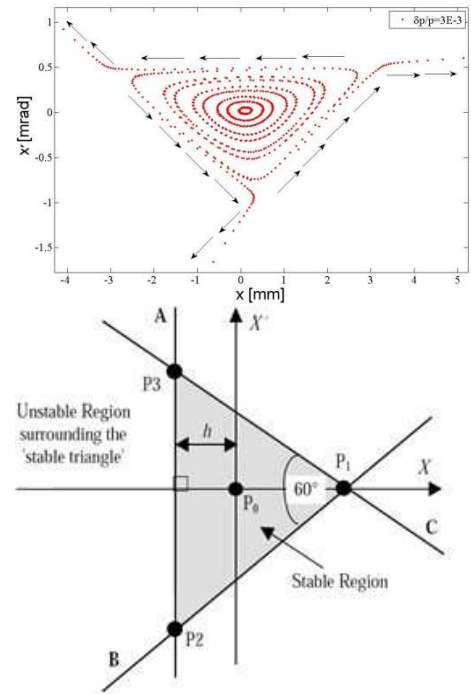

**Fig. 7.** Diagram of the third order resonance. The upper and the bottom represent numerical tracking and theoretical analysis respectively.

Table 2: The distance $h$ comparison between the theoretical value and numerical tracking

|  | 1 | 2 | 3 | 4 | 5 | 6 |
|---|---|---|---|---|---|---|
| Chromatic sext. | S | S | S | S | S | S |
| Harmonic sext. | 0 | -S | -2S | -3S | -4S | -5S |
| Theoretical $h$ [*10$^4$m$^{1/2}$] | 34.8 | 17.9 | 11.6 | 8.70 | 6.96 | 5.8 |
| Numerical $h$ [*10$^4$m$^{1/2}$] | 32.1 | 16.3 | 10.8 | 7.92 | 6.50 | 5.33 |

### b)  RF knock-out method for slow extraction

RF knock-out method can provide high irradiation accuracy even for an irregular target, and result in high beam-utilization efficiency. And also RF knock-out method is useful in the spot scanning method, because the beam supply can be easily started or stopped in the extraction period [8]. RF kicker is a dipole which can provide the transverse kick with related to the beam tune $Q_x$. After the separatrix is fixed, the RF knock-out method can make the beam emmittance growth and then the beam can be extracted. Figure 8 shows the position of a particle versus the revolution turn. By tracking the position of particle with RF kicker, we found the position of the particle grows linearly around 2000 turns, and decreases linearly in a symmetrical pattern, and then repeats the above process.

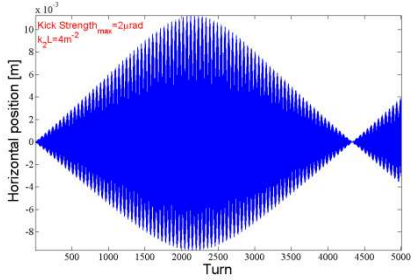

**Fig. 8.** The position of the particles versus revolution turns with the effects of the RF kicker in a mono frequency.

For a Gaussian distribution beam, if we suppose the distributions are independent and identically in X direction and X' direction, the module of the position and velocity can be expressed by Rayleigh distribution [9] in radial direction as $p(r) = \frac{2r}{\sigma^2}\exp[-\frac{r^2}{\sigma^2}]$, where $\sigma$ is the standard deviation of the Gaussian distribution, and $r$ corresponds to the radial length. Given a Gaussian distribution beam contains 1000 particles, a RF kicker is used to excite the beam in 1000 turns, and the number of extracted particles is 652, and the spill structure is depicted in figure 9. By using a Rayleigh distribution function to fill the spill, the result is agreed well.

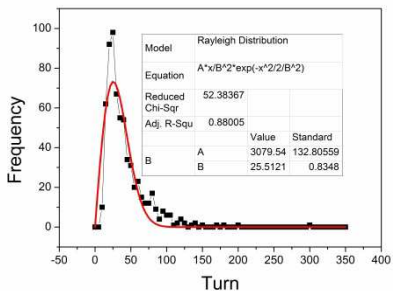

**Fig. 9.** The spill character extracted from a Gaussian distribution beam.

### c) slow beam extraction by RF field with amplitude modulation

From the above section, we found that the extracted beam can be expressed as a Rayleigh distribution for a Gaussian distribution beam. After a lot of simulations, we are excited to find that for a mono frequency RF kicker, the spill character is related to the RF kicker's amplitude. So getting a more suitable amplitude curve of the RF kicker, also called amplitude modulation, is vital important in the whole extraction period. Now a simple model is used to get the amplitude modulation curve of the transverse RF kicker [9].

The number of the extracted particles, $N_{ext}$ can be expressed as

$$N_{ext} = \int_{r_0}^{\infty} N_0 p(r)dr = N_0 \exp[-\frac{r_0^2}{\sigma^2(n)}], \quad (1)$$

where $N_0$ is the total number of the particles, and $\sigma$ is the standard deviation of the Rayleigh distribution. Thus, the time structure of the beam can be represent as

$$\frac{dN_{ext}}{dn} = N_0 \frac{d\sigma^2(n)}{dn}\frac{r_0^2}{\sigma^4(n)}\exp(-\frac{r_0^2}{\sigma^2(n)}). \quad (2)$$

Considering beam dilution by RF kicker, the relationship between $\sigma(n)$ and $\theta(n)$ can be expressed as

$$\sigma^2(n) = k\theta(n)^2 n + \sigma_0^2, \quad (3)$$

where $\sigma_0$ is the initial standard deviation value, $k$ is a constant number related to beam distribution, and $\theta(n)$ is the kicker amplitude in the $n^{th}$ turn.

The rate of the beam extraction can also be expressed as

$$\frac{dN_{ext}}{dn} = \frac{N_0}{f_{rev}\tau_{ext}}\left\{1-\exp(-\frac{r_0^2}{\sigma_0^2})\right\}, \quad (4)$$

where $f_{rev}$ is the frequency of the particles, and $\tau_{ext}$ is the total extraction time. In order to keep equation 4 constant, by using equation 2 and equation 3, the RF kicker amplitude can be figured as

$$\theta(n) = [\frac{d}{dn}(\frac{\sigma^2(n)}{k})]^{1/2}, \quad (5)$$

where

$$\sigma^2(n) = -r_0^2[\ln[\frac{n}{f_{rev}\tau_{ext}}\{1-\exp(-\frac{r_0^2}{\sigma_0^2})\}+\exp(-\frac{r_0^2}{\sigma_0^2})]]^{-1}. \quad (6)$$

By using $k = 312$ that can be figured by fitting constant RF kicker amplitude with different values, $r_0 = 9 \times 10^{-3}$, and $\sigma_0 = 3.2 \times 10^{-3}$, $\tau_{ext} = 1.6s$, the relationship between RF kicker amplitude versus revolution number can be described as figure 10.

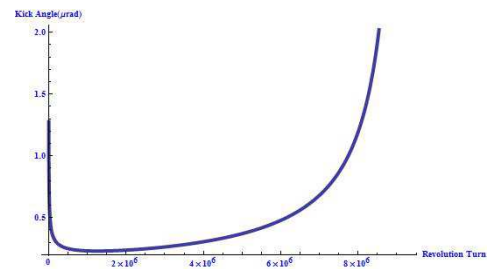

**Fig. 10.** The RF kicker amplitude versus revolution number.

## 4  Summary and discussion

A radiation synchrotron with circumference of 33.6 has been designed. A multi-turn injection method with 2-bump is adopted for beam injection. And the injection bump height is carefully optimised to make the distribution of the injection beam more uniform in normalised phase space. For considering saving the strength the sextuples, the combination of two sextuples with different strength whose phase advance is $\pi/3$ has been carefully studied, and one can conclude that sextuples with $\pi/3$ phase advance with the opposite polarity enforce the driving term of the third order resonance, and it is very useful for the sextuple planning in a small proton accelerators. The numbers of the extracted particles distributed in the Rayleigh distribution have been proven and a RF knock-out method with amplitude modulation is used to smooth the spill.

*We wish to thank Zhang Manzhou in SSRF for his helpful and useful discussions on third order resonance.*